\documentclass[]{tMPH2e}
\usepackage{longtable}
\usepackage{rotating}
\usepackage{lscape}

\begin{document}

\newcommand{\wn}{~\ensuremath{\mathrm{cm^{-1}}}}
\newcommand{\etal}{\emph{et al.}}
\newcommand{\dstate}{$D^{1}\Pi_{u}$}
\newcommand{\xstate}{$X^{1}\Sigma_{g}^{+}$}
\newcommand{\pmin}{$\Pi^{-}$}
\newcommand{\pplus}{$\Pi^{+}$}

\title{VUV Spectroscopic Study of the \dstate\ State of Molecular Deuterium}

\author{G.D. Dickenson$^{a}$, T.I. Ivanov$^{a}$,  W. Ubachs$^{a}$ $^{\ast}$\thanks{$^\ast$Corresponding author. Email: w.m.g.ubachs@vu.nl}, M. Roudjane$^{bd}$ ${^\dagger}$\thanks{$^\dagger$present adress : Department of Chemistry, 009 Chemistry-Physics Building, University of Kentucky, 505 Rose Street, Lexington, KY 40506-0055, USA}, N. de Oliveira$^{b}$, D. Joyeux$^{bc}$, L. Nahon$^{b}$, W.-\"{U} L. Tchang-Brillet$^{de}$ \\ \vspace{10pt} M. Glass-Maujean$^{f}$, H. Schmoranzer$^{g}$, A. Knie$^{h}$, S. K\"{u}bler$^{h}$ and A. Ehresmann$^{h}$}

\maketitle
\begin{center}
\begin{scriptsize}
$^{a}$Institute for Lasers, Life and Biophotonics, VU University, De Boelelaan 1081, 1081 HV Amsterdam, The Netherlands\\
$^{b}$Synchrotron Soleil, Orme des Merisiers, St Aubin BP 48, 91192 GIF sur Yvette cedex, France\\
$^{c}$Laboratoire Charles Fabry de l'Institut d'Optique, CNRS, Univ Paris-Sud Campus Polytechnique, RD128, 91127, Palaiseau cedex, France\\
$^{d}$Laboratoire d'\'Etude du Rayonnement et de la Mati\`ere en Astrophysique, UMR 8112 du CNRS, Observatoire de Paris-Meudon, 5 place Jules Janssen, 92195 Meudon cedex, France\\
$^{e}$Universit\'e Pierre et Marie Curie-Paris 6\\
$^{f}$Laboratoire de Physique Mol\'eculaire pour l'Atmosph\`ere et l'Astrophysique UMR7092, Universit\'e P. et M. Curie, 4 place Jussieu 75252 Paris cedex 05 France\\
$^{g}$Fachbereich Physik, Technische Universit\"{a}t Kaiserslautern, D-67653 Kaiserslautern\\
$^{h}$Institut f\"{u}r Physik, Universit\"{a}t Kassel, D-34132 Kassel, Germany\\
\end{scriptsize}
\end{center}
\vspace{15pt}

\begin{abstract}

The $D^{1}\Pi_{u}$ - $X^{1}\Sigma_{g}^{+}$ absorption system of molecular deuterium has been re-investigated using the VUV Fourier -Transform (FT) spectrometer at the DESIRS beamline of the synchrotron SOLEIL and photon-induced fluorescence spectrometry (PIFS) using the 10 m normal incidence monochromator at the synchrotron BESSY II. Using the FT spectrometer absorption spectra in the range 72 - 82 nm were recorded in quasi static gas at 100 K and in a free flowing jet at a spectroscopic resolution of 0.50 and 0.20 \wn\ respectively . The narrow Q-branch transitions, probing states of \pmin\ symmetry, were observed up to vibrational level $v$ = 22.  The states of \pplus\ symmetry, known to be broadened due to predissociation and giving rise to asymmetric Beutler-Fano resonances, were studied up to $v$ = 18.
The 10 m normal incidence beamline setup at BESSY II was used to simultaneously record absorption, dissociation, ionization and fluorescence decay channels from which information on the line intensities, predissociated widths, and Fano $q$-parameters were extracted.  R-branch transitions were observed up to $v$ = 23 for $J$ = 1-3 as well as several transitions for $J$ = 4 and 5 up to $v$ = 22 and 18 respectively. The Q-branch transitions are found to weakly predissociate and were observed from $v$ = 8 to the final vibrational level of the state $v$ = 23. The spectroscopic study is supported by two theoretical frameworks. Results on the \pmin\ symmetry states are compared to \emph{ab initio} multi-channel-quantum defect theory (MQDT) calculations, demonstrating that these calculations are accurate to within 0.5 \wn . Furthermore, the calculated line intensities of Q-lines agree well with measured values. For the states of \pplus\ symmetry a perturbative model based on a single bound state interacting with a predissociation continuum was explored, yielding good agreement for predissociation widths, Fano $q$-parameters and line intensities.

\end{abstract}

\section{Introduction}

As the simplest neutral molecule hydrogen plays an important role as a testing ground for fundamental quantum theory calculations. The extension to molecular deuterium provides information on mass-dependent effects in the molecular level structure. In particular the $D^{1}\Pi_{u}^{+}$ state of molecular hydrogen is known to predissociate and is the benchmark case of this phenomenon in neutral molecules. A recent investigation on this predissociation phenomenon for H$_{2}$~\cite{Dickenson2010a} is now extended to D$_{2}$. Spectral measurements of D$_{2}$ also have practical applications; D$_{2}$ has been detected in its excited state in cold regions of tokamaks and plays a role in radiative losses in fusion plasmas~\cite{Fantz1998}.

The first predissociated level in the $D^{1}\Pi_{u}^{+}$ state is $v = 4$; for $v\leq 3$ levels, lying below the $n$ = 2 dissociation threshold, a sequence of sharp resonances is found. Beutler and coworkers were the first to observe this in the $D^{1}\Pi_{u}^{+}$  - \xstate\ system in D$_{2}$ (and in H$_{2}$, for which $v$ = 3 is the first predissociated level) and correctly assigned the sequence of vibrational levels up to $v$ = 7~\cite{Beutler1935a}.  Higher resolution studies, conducted by Monfils~\cite{Monfils1965}, confirmed these vibrational assignments and extended them up to $v$ = 15 and to high rotational states ($J$ up to 9), for both $\Pi^+$ and $\Pi^-$ parity components. 

The first study of the predissociated lineshapes, based on the theory of Fano~\cite{Fano1935,Fano1961} for H$_{2}$ and D$_{2}$ was reported by Comes and Schumpe~\cite{Comes1971}.  Calculations by Fiquet-Fayard and Gallais and Jullienne~\cite{Fiquet-Fayard1971, Julienne1971} were performed in the early seventies. A discrepancy between experiment and theory concerning the predissociated widths for H$_{2}$ was resolved by Fiquet-Fayard and Gallais~\cite{Fiquet-Fayard1972}. Further classical absorption studies conducted on the predissociated widths of D$_{2}$ were made by Dehmer and Chupka~\cite{Dehmer1980}. Laser based studies were conducted by Rothschild \etal\ ~\cite{Rothschild1981}, employing a laser with bandwidth 0.005 \wn\ yielding accurate information on the predissociated widths for the $D^{1}\Pi_{u}^{+}$  $v = 5$ state. 

The $\Pi_u^-$ symmetry components of the $D^1\Pi_u$ system has been studied by emission in the VUV range by Larzilli\`ere \etal~\cite{Larzilliere1980} for excited state vibrations $v=5-12$.   An atlas of the visible emission measurements of G.H. Dieke containing over 27 488 lines of deuterium was compiled by Freund \etal~\cite{Freund1985}; in it level energies for \pplus\ parity components of the \dstate\ state were determined for the unpredissociated levels $v=0-3$ while the \pmin\ parity components were listed up to $v=12$. More recently, improved, highly accurate emission studies, again in the VUV range, were conducted by Roudjane \etal\ ~\cite{Roudjane2006}, in which level energies for the \pmin\ states were determined up until $v=19$ at an accuracy of 0.3 \wn .

We present extensive measurements on the \dstate\ state of D$_{2}$ using both the FT-spectrometer located on the DESIRS beamline of the synchrotron SOLEIL and a setup for combined absorption, ionization yield and fluorescence measurements~\cite{Reichardt2001} located on the U125/2 undulator beamline of the synchrotron BESSY II. The two experimental techniques are complementary and an analysis of both yields an almost complete characterization of the \dstate\ state of D$_{2}$. The FT-spectrometer extends the technique of FT spectroscopy down to vacuum ultraviolet wavelengths of 40 nm without the use of a beam splitter~\cite{deOliveira2009,deOliveira2011}. Line positions and widths were determined for the predissociated \pplus\ components up to $v=18$ for $J=1$ and 2 and for the \pmin\ components line positions were determined up to vibrational level $v=22$ for $J=1$, to an accuracy of 0.06 \wn .

The setup at BESSY II has the capability of measuring simultaneously absorption, dissociation, ionization and molecular fluorescence, even to the extent to unravel the competing decay channels and to derive absolute cross sections. After studies in H$_{2}$ ~\cite{Glass-Maujean2010,Glass-Maujean2010a} now the \dstate\ state of D$_{2}$ has been investigated. Predissociated widths were extracted from these measurements and compared to those from the absorption data taken at SOLEIL for $J=1$ and 2 for $v=8-18$. Furthermore predissociated widths, intensities and $q$-parameters for transitions probing rotational levels $J=3,4$ and 5 were extracted from these spectra and compared to theory. For the $\Pi^{-}$ components line intensities were determined up to vibrational level $v=18$ for $J=1-4$ and up to 22 for $J=2$. In addition the line positions were obtained up to the last vibrational level $v=23$.

Two different theoretical approaches are adopted to interpret the experimental results. For levels of \pplus\ symmetry, line widths and shapes are calculated applying Fano’s perturbative method for a single bound state interacting with a single continuum~\cite{Fano1935,Fano1961} . This calculation has the same starting point as the ones by Fiquet-Fayard and Gallais~\cite{Fiquet-Fayard1971} and Jullienne~\cite{Julienne1971}, focussing on the predissociated widths of the $D^{1}\Pi^{+}$ components, but extended to D$_{2}$ and using updated \emph{ab-initio} potentials and coupling functions~\cite{Wolniewicz2003}. The predissociated widths, Fano-$q$-parameters and intensities have been calculated for the R-branch of the $D-X$ system up to $v=23$ for $J=1-5$ using this perturbative calculation.

For the narrow resonances probing the \pmin\ levels calculations are performed in a more refined multi-channel quantum defect (MQDT) approach, including non-adiabatic interactions. This model, originally developed by Jungen and Atabek~\cite{Jungen1977}, was recently worked out in detail to interpret intensities for Q-branch lines in the $D-X$ system of H$_{2}$ and D$_{2}$~\cite{Glass-Maujean2009,Glass-Maujean2007,Glass-Maujean2011}. Subsequently this model was used to interpret the competing decay channels (fluorescence, dissociation and ionization) in the $D^{1}\Pi_{u}^{-}$  state of H$_{2}$~\cite{Glass-Maujean2010,Glass-Maujean2010a}.

\section{Experimental Setup}

The absorption measurements analysed here, taken with the FT-spectrometer, were recorded in two different configurations of the interaction chamber, a windowless cell-configuration and a free jet-configuration. Only the most important aspects are highlighted. For a detailed description of the FT-spectrometer we refer to Refs.~\cite{deOliveira2009,Ivanov2010,deOliveira2011}. For the broad, predissociated transitions ($v\geqslant 4$) the cell-configuration was used. The synchrotron radiation passes through a cell of quasi-static D$_{2}$, approximately 10 cm long, before entering the FT-spectrometer. The cell is enclosed by a cylinder through which liquid nitrogen is allowed to flow. This cools the cell down to approximately 100 K, reducing the Doppler width and making the spectra less congested. The time constraints of a synchrotron run requires a compromise between the signal to noise ratio, the targeted resolution and the spectral window. Considering the expected Doppler broadening at 100 K ($\sim$0.44 \wn ), the resolution linewidth of the spectrometer was set to 0.35 \wn .

A molecular jet-configuration, operated continuously, was used to measure the narrow unpredissociated R-branch transitions ($v\leq 3$) and several Q-branch transitions ($v\leq$ 7). The synchrotron radiation passes through a free-flowing jet of D$_{2}$ thus reducing Doppler broadening. For this particular configuration, the FTS resolution was set to 0.16 \wn . Due to insufficient differential pumping efficiency, the remaining room-temperature background gas contributes as a broad pedestal underlying the narrow Doppler-free feature of the line shape. The narrow feature at a spectral width of 0.2 \wn\ is deconvolved from this composite line shape and used for calibrating the resonances.

\begin{figure}
\centering
\includegraphics[width=0.8\linewidth]{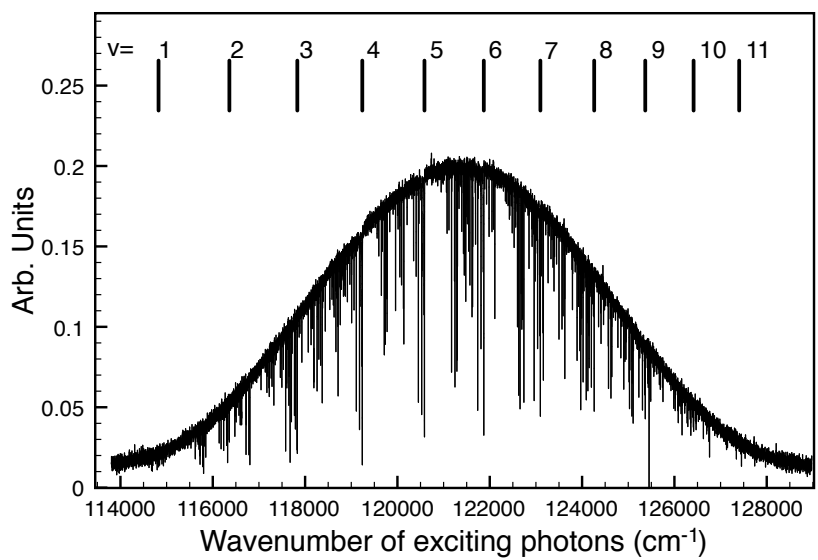}
\caption{A typical undulator output profile of the DESIRS beamline of the synchrotron SOLEIL. Several absorption bands of the \dstate\ - $X^{1}\Sigma_{g}^{+}(v',0)$ state are indicated.}
\label{fig:BellCurve}
\end{figure}

The bell-shaped output of the undulator radiation at the DESIRS beamline of the synchrotron SOLEIL, illustrated in Fig.~\ref{fig:BellCurve}, spans approximately 5000 \wn . The measurements were split into four overlapping windows spanning 115 000 - 134 000 \wn . Each window, averaged over 100 interferograms taking approximately 2 hours to accumulate, was measured at several pressures with the aim of recording all lines under unsaturated conditions. 

The spectra recorded at the BESSY II Synchrotron in Berlin were measured with a 10 m normal incidence monochromator at the undulator beamline U125/2 equipped with a 4800 lines/mm grating yielding a spectral resolution of 1.6-2.5 \wn\ \cite{Glass-Maujean2007a}. The differentially-pumped absorption cell is 39 mm long and contains 26.7 $\mu$bar of D$ _{2}$ at room temperature. The transmitted light intensity needed for determining the absorption cross section was detected by a photodiode located at the back of the cell. Molecular visible fluorescence and Ly-$\alpha$ fluorescence radiation from the D($n=2$) atomic fragments were recorded with a visible sensitive detector and a VUV photomultiplier respectively~\cite{Glass-Maujean2010}. Both detectors are placed at the magic angle with respect to the polarization direction of the incident light, thereby avoiding any polarization effects. In addition, photoions were collected by applying a voltage of 10 V to one of two electrodes in the target cell~\cite{Glass-Maujean2010}. All spectra were recorded simultaneously as a function of incident photon energy, while the measurements cover the range 124 000 - 136 000 \wn . The setup and measurement procedures have been refined to extract the yields of the competing decay channels; fluorescence, dissociation and ionization~\cite{Glass-Maujean2010a}. An analysis of the dissociation excitation spectrum, henceforth referred to as the Ly$-\alpha$ spectrum, yields information on intensities, predissociated widths and Fano $q$-parameters of the transitions to the $D^{1}\Pi_{u}^{+}$ state of D$_{2}$.

\begin{figure}[b]
\begin{center}
\includegraphics[width=0.7\linewidth]{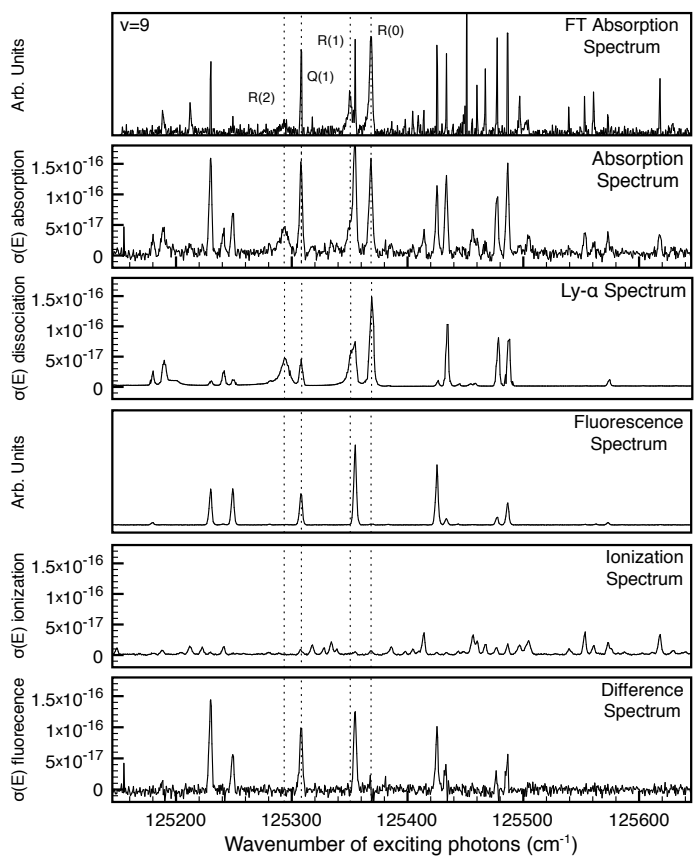}
\caption{A portion of the measured spectra analysed in the present study. The FT absorption spectrum measured at SOLEIL, presented at the top, compared to the spectra measured at BESSY II (absorption, Ly-$\alpha$, fluorescence, ionization and difference). The y-axis of the FT absorption spectrum and fluorescence spectrum have arbitrary units whereas for the absorption, Ly-$\alpha$, ionization and difference spectrum the y-axis represents cross section measurements in cm$^{2}$. Different aspects of the line profiles are extracted from the different spectra (see text for details). The R(0),R(1),Q(1) and R(2) transitions of the (9,0) band are indicated.}
\label{fig:SpectralComparison}
\end{center}
\end{figure}

Figure~\ref{fig:SpectralComparison} shows a comparison between the FT absorption spectra (obtained at SOLEIL) and the absorption, Ly-$\alpha$, fluorescence, ionization and difference spectra (see section~\ref{sec:ExpResAndDis}) recorded at BESSY II over $\sim$500 \wn . The figure illustrates the advantages of the two experimental techniques. The FT absorption spectrum has superior resolution which allows for an accurate determination of the line positions, whereas the spectra recorded with the monochromator are measured on a zero background, thus facilitating the profile analysis. It should be emphasised that the base lines of the Ly-$\alpha$ and ionization spectra are not the noise level but are due to the dissociation and ionization continua respectively. Furthermore the spectra at BESSY II yield information on absolute cross sections and on rates of competing decay channels.

\section{Theory}
\subsection{Perturbative Calculations for the $D^{1}\Pi^{+}_{u}$ levels}
\label{sec:Calculation}
The theoretical approach for the calculation of the $D^{1}\Pi^{+}_{u}$ levels is based upon Fano's description of a bound state interacting with a continuum~\cite{Fano1935,Fano1961}. Fiquet-Fayard and Gallais~\cite{Fiquet-Fayard1971} have taken this approach before. These calculations have been improved by including new accurate $D$ and $B'$ potentials, transition moments and coupling functions~\cite{Wolniewicz2003}. We provide a brief description of the details of the calculation.

\begin{figure}
\centering
\includegraphics[width=0.8\linewidth]{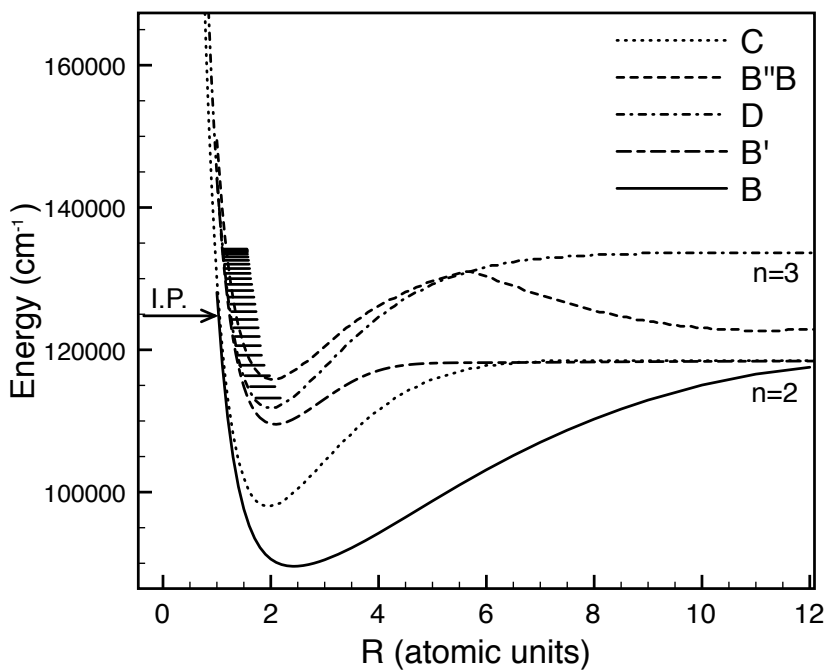}
\caption{The potential energy curves of the relevant excited electronic states~\cite{Dressler1986}. The vibrational levels in the $3p\pi D^{1}\Pi_{u}$ state are indicated as well as the ionization potential for D$_2$~\cite{Liu2010}.}
\label{fig:PotCurves}
\end{figure}

The strong Coriolis coupling to the continuum of the $B'^{1}\Sigma_{u}^{+}$ state is the dominant cause of the predissociation in the $ D^{1}\Pi^{+}_{u}$ levels~\cite{Monfils1961,Julienne1971,Fiquet-Fayard1971} while the continuum of the $C^{1}\Pi_{u}$ state contributes only marginally to the predissociation. $ D^{1}\Pi^{-} $ levels do not couple to the $B'^{1}\Sigma_{u}^{+}$ continuum due to symmetry reasons. They do however predissociate very slightly as a direct consequence of the coupling to the $C^{1}\Pi_{u}^{-}$ continuum~\cite{Glass-Maujean2010a,Glass-Maujean2010}. The potential energy curves~\cite{Dressler1986} for the aforementioned electronic states are illustrated in Fig.~\ref{fig:PotCurves} along with the curve of the $B''\bar{B}^{1}\Sigma^{+}_{u}$ state~\cite{deLanga2001}, which is also predissociated due to a coupling to the $B'$ state~\cite{Glass-Maujean2007}.

The predissociation produces typical Beutler-Fano resonances which have the form
\begin{eqnarray}
\dfrac{(q+\epsilon)^{2}}{1+\epsilon^{2}}
\label{eqn:Fano} 
\end{eqnarray}
where 
\begin{eqnarray}
\epsilon = \dfrac{E-E_{0}}{\Gamma_{v}}
\label{eqn:epsilon} 
\end{eqnarray}
and $q$ describes the asymmetry in the absorption profile induced by the predissociation,
\begin{eqnarray}
   q = \frac{<\varphi_{Dv'}|T_{DX}(R)|\varphi_{Xv''}>}{\pi V<\varphi_{B'\epsilon}|T_{B'X}(R)|\varphi_{Xv''}>}
\label{eqn:qfactor}
\end{eqnarray}
with transition strength matrix elements in the nominator (discrete state excitation) and denominator (continuum excitation) and
\begin{eqnarray}
 V= \langle\varphi_{B'\varepsilon}|H(R)|\varphi_{Dv'}\rangle
\label{Vfactor}
\end{eqnarray}
representing the interaction matrix element between the discrete $D$ state and the $B'$ continuum. The operator $H(R)$ is:
\begin{eqnarray}
H(R) =\dfrac{m}{\mu} \sqrt{\dfrac{J(J+1)}{2}}\dfrac{<\psi_{B'}|L^{+}|\psi_{D}>}{R^{2}}
\label{eqn:Proportionality}
\end{eqnarray}
where $m$ is the mass of the electron, $\mu$ is the reduced mass of the molecule, $J$ is the rotational quantum number (excited state) and $\psi_{B'}$ and $\psi_{D}$ are the electronic wave functions of the $B'$ and $D$ states respectively~\cite{Wolniewicz2006}.
 The widths $\Gamma_{v}$ derive from lifetime shortening due to predissociation and are related to the $D$-$B'$ interaction through the square of the interaction matrix element: 
\begin{eqnarray}
\Gamma_{v}=2\pi V^{2} \propto J(J+1)
\label{eqn:Width}
\end{eqnarray}
From Eq.(\ref{eqn:Proportionality}) it follows that the predissociation widths scale with $\mu^{-2}$, resulting in smaller values for D$ _{2} $ compared to H$ _{2} $ by a factor 4. The eigenfunctions used in Eq.(\ref{eqn:qfactor}) and (\ref{eqn:Proportionality}) were evaluated through a Numerov integration of the Schr\"{o}dinger equation applied to the adiabatic potentials of Wolniewicz and Staszewska~\cite{Wolniewicz2003} ($D$ state) and Staszewska and Wolniewicz~\cite{Staszewska2002} ($B'$ state). The ($D-X$) transition  moment functions of Eq.(\ref{eqn:qfactor}) are from Ref.~\cite{Wolniewicz2003} and the $B'-X$ transitions moments are from Ref.~\cite{Wolniewicz2003a} and the matrix element of Eq.(\ref{eqn:Proportionality}) is obtained from Ref.~\cite{Wolniewicz2006}.

The Einstein $A$ coefficients can be calculated from the matrix elements through the relation~\cite{Glass-Maujean2010}:
\begin{eqnarray}
A_{n\rightarrow v',v''} = \dfrac{4}{3} \dfrac{mc^{2}}{\hbar}\alpha^{5}(\dfrac{E_{n}-E_{v''N''}}{2Rhc})\vert \dfrac{\langle \varphi_{Dv'}|T_{DX}(R)|\varphi_{Xv''}\rangle}{a_{0}}\vert ^{2}
\label{eqn:A coefficient}
\end{eqnarray}
where $\alpha$ is the fine structure constant, $m$ is the mass of the electron, $R$ is the Rydberg constant and $a_{0}$ is the Bohr radius. The ratios in brackets $(...)$ and $\vert ... \vert $ correspond to the transition energy and the dipole transition moment in atomic units. The integrated absorption cross sections (line intensities) for a discrete level are related to the Einstein $A$ coefficients (see sec.~\ref{sec:Intensities}).

Hence, this perturbative approach provides the position and intensities of the lines as well as the $q$ and $\Gamma_{v}$ line shape parameters for the \pplus\ symmetry levels investigated in the present study. It is noted that the ionization channel (occuring for levels $v> 8$) as well as other non-adiabatic couplings are neglected in this approach.

\subsection{Non-adiabatic MQDT calculation for $D^{1}\Pi_{u}^{-}$ levels}
\label{sec:CalculationMQDT}

The framework of \emph{ab-initio} multichannel quantum defect (MQDT) calculations for the states of ungerade symmetry in molecular hydrogen dates back to the pivotal paper by Jungen and Atabek~\cite{Jungen1977}. Calculations focussing on the levels of $\Pi_{u}^{-}$ symmetry were carried out to interpret synchrotron data of the Q-branch lines in H$_{2}$~\cite{Glass-Maujean2009}. These calculations have been extended to D$_{2}$ in particular to test the MQDT for predicting line positions and intensities of narrow super-excited states~\cite{Glass-Maujean2011}.  In Ref.~\cite{Glass-Maujean2011}, these calculations are compared with calculations carried out in Ref.~\cite{Roudjane2006} by solving coupled differential equations for the four electronic states $B, C, D$ and $B'$, which take into account their non-adiabatic interactions. 

\section{Experimental Results and Discussion}
\label{sec:ExpResAndDis}
Accurate line positions and widths of the $D-X$ resonances are obtained from the FT absorption spectra by fitting the data with a model that correctly describes the recorded convolved profiles. In a first step the bell-shaped background continuum of the undulator profile (see Fig.~\ref{fig:BellCurve}) of the FT absorption measurements was fitted to a spline and divided through resulting in a flat continuum. The transitions were then fitted with a model constructed by first convolving the line profile (either Lorentzian for the non-predissociated states or Beutler-Fano for the predissociated states) with a Gaussian function representing the Doppler width. The non-linearity in the absorption depth due to the Beer-Lambert law is included in the model and finally the profile is convolved with the instrument function which is a sinc function. The model was then fitted to the data using a standard least-squares optimization routine. 

\begin{figure}[t]
\begin{center}
\includegraphics[width=0.7\linewidth]{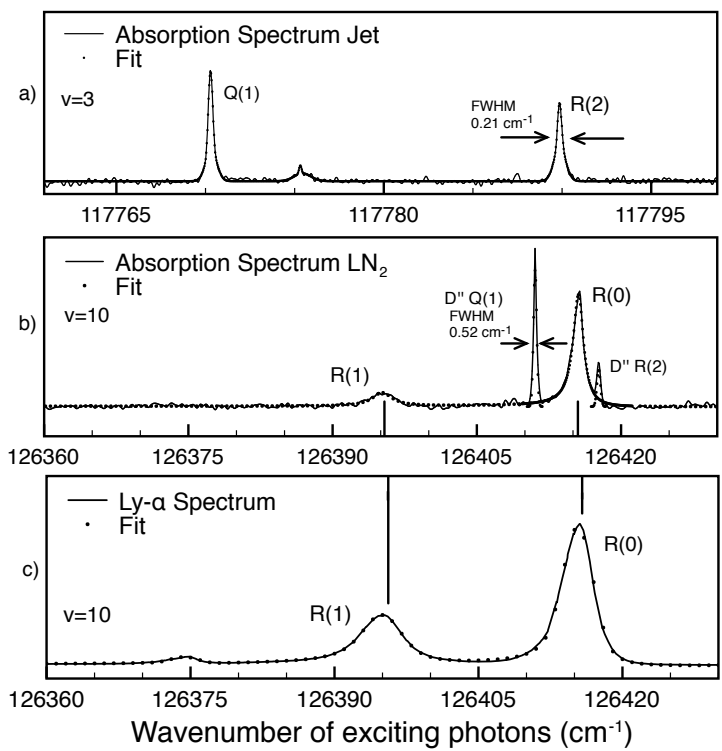}
\caption{ Recorded spectra (solid lines) of some $D - X$($v'$,0) bands using the VUV FT spectrometer at SOLEIL and the scanning monochromator at BESSY II. (a) The R(2) and Q(1) line of the (3,0) band, as recorded in the jet configuration of the FT spectrometer at a FWHM of 0.21 \wn , is illustrated along with the resulting fit (dotted line). (b) the (10,0) band, recorded with the FT spectrometer in the 100 K static cell configuration, shows typical Beutler-Fano line shapes due to predissociation. The unpredissociated Q(1) and R(2) transitions of the $D''-X$ (4,0) band are indicated at a FWHM of 0.52 \wn . (c) the (10,0) band recorded using the BESSY II scanning monochromator; note that the two transitions overlapped with the R(0) line in the absorption spectrum are not present here as they do not predissociate.}
\label{fig:SoleilVsBessy}
\end{center}
\end{figure}

Small sections of the spectrum are depicted in Fig. \ref{fig:SoleilVsBessy} along with their corresponding fit. Fig. \ref{fig:SoleilVsBessy}(a) depicts two Doppler-broadened transitions from the (3,0) band (R(2) and Q(1)) probing unpredissociated states below the second dissociation limit recorded in the jet-configuration. In order to estimate the spectral resolution in this configuration, each transition was fitted with two Gaussian functions,  representing the room temperature pedestal due to background gas and the narrow feature due to the jet. The width of the narrow feature of the R(2) line,  $\sim$ 0.21 \wn , serves to illustrate the limiting spectral resolution in the jet-configuration. Figure~\ref{fig:SoleilVsBessy} (b) shows the $D-X$ (10,0), R(0) and R(1) transitions and the $D''-X$ (4,0) Q(1) and R(2) transitions~\cite{Takezawa1975} recorded with the cell-configuration. The FWHM of the $D''-X$ Q(2) transition, fitted with a single Gaussian function is approximately 0.5 \wn , serves to illustrate the limiting spectral resolution of this configuration.

FT spectra have an intrinsic calibration, providing a linear frequency scale.  A single atomic Ar line $(3p)^{5}(^{2}$P$_{3/2})9d \leftarrow (3p)^{6}$ $^{1}$S$_{0}$, known to an accuracy of 0.03 \wn\ was used as the reference line for an absolute calibration~\cite{Sommavilla2002}. The ultimate precision of the FT-spectrometer with the present settings was determined in a previous study~\cite{Ivanov2010} to be 0.04 \wn . This precision is dependent on the signal-to-noise ratio and the number of measurement points the line shape consists of. We estimate an uncertainty of 0.06 \wn\ for the narrow transitions in the Q branch for $v=4-18$ and for the transitions in the unpredissociated bands. For the slightly saturated Q-branch transitions the uncertainty is estimated at 0.08 \wn . For the  measured predissociated transitions probing \pplus\ levels $v=4-18$, $J=1,2$ the estimated uncertainty is 0.2 \wn . For strongly blended lines (discussed in section~\ref{sec:SpectroscopicAnalysis:Absorption Spectrum}), saturated lines and weaker lines probing the rotational level $J=3$ the uncertainty is estimated at 0.4 \wn .

The absolute calibration of the frequency scale of the spectra obtained at BESSY II, is derived from the FT-VUV data obtained at SOLEIL. The linearity of the frequency scale, which was recently improved by the Heydemann correction~\cite{Follath2010}, remains the main source of uncertainty on the energies of the measured spectra. From the comparison with the FT measured positions it can be confirmed that the uncertainty on these values is typically $\pm$ 1.0 \wn . The cross sections of the BESSY II absorption spectrum have been calibrated directly, based on the known pressure and the absorption path length. In order to evaluate the importance of the continua, certain points have been recorded without gas at the beginning and at the end of the scanning procedure. The sources of uncertainty in the absolute cross-section measurements (length, pressure and temperature) are estimated to a total error of 6$\%$, which has to be added to the statistical error due to the noise at each recorded point ~\cite{Glass-Maujean2007}.
The Ly-$\alpha$ and ion detection efficiencies depend on the geometry and the detectors used, but are independent of the wavelength. The detection efficiencies are determined by comparing absorption and dissociation or ionization structures in the spectrum for lines that have a 100$\%$ dissociation or ionization yield. The uncertainty of the dissociation or ionization yields thus obtained is estimated to be 5$\%$, which has to be added to the 6$\%$ error of calibration and to the error due to the noise at each recorded point. The $D^{1}\Pi^{+}$ levels with $v>3$ are absent in the ionization spectrum since the ionization and radiation rates for these levels are below our detection capability. For this specific case the dissociation yield is considered to be 100$\%$ and the Ly-$\alpha$ spectrum reproduces the absorption spectrum for the R and P lines belonging to the $D-X$ system.
The fluorescence signal has been recorded undispersedly by a visible sensitive photomultiplier. Therefore this signal cannot be calibrated to absolute cross sections, since the spectral quantum efficiency of the multiplier varies with varying wavelengths, i.e. from one level to the other. Quantitative information on the fluorescence cross sections is obtained by subtracting the dissociation and ionization spectra from the absorption spectrum~\cite{Glass-Maujean2007}
\begin{eqnarray}
\sigma_{fluo}=\sigma_{dif}=\sigma_{abs}-\sigma_{diss}-\sigma_{ion}
\label{eqn:FluorescenceSpectrum}
\end{eqnarray}
and is referred to as the difference spectrum. The $D^{1}\Pi^{-}_{u}$ levels radiate and therefore appear in the fluorescence spectrum, which is less congested and has a better signal to noise ratio than the absorption spectrum. Their dissociation and ionization rates are below 5$\%$ and they are hardly visible in the dissociation or ionization spectra. The difference spectrum thus reproduces the absorption spectrum for these transitions, within an uncertainty of 15-20$\%$ and represents a relatively sparse spectrum from which several absolute cross sections are determined for the $D^{1}\Pi^{-}_{u}$ states~\cite{Glass-Maujean2010a}. 

Figure~\ref{fig:SoleilVsBessy}(b) and (c) show a comparison between the FT absorption and Ly-$\alpha$ spectra for $v=10$. The widths of the Beutler-Fano resonances appear larger in panel c)  due to the larger instrumental width of the scanning monochromator ~2 \wn\ compared to 0.35 \wn\ of the FT-spectrometer. To illustrate an advantage of the BESSY II spectra the Q(1) and R(2) transitions from the (4,0) band of the $D''-X$ system~\cite{Takezawa1975} obscuring the R(0) transition are not present in the Ly-$\alpha$ spectrum since these states do not predissociate. The measured line positions indicated by a solid vertical line are deconvolved from the asymmetric line profiles. Note that in the case of a Beutler-Fano profile the deduced transition frequency does not coincide with the peak of the spectral line.

\subsection{Spectroscopic Analysis: FT Absorption Spectrum}
\label{sec:SpectroscopicAnalysis:Absorption Spectrum}
The measurements of the transition frequencies for the Q branch transitions probing \pmin\ levels are listed in Tab.~\ref{Tab:Pi-}  and those of the R and P branches probing \pplus\ levels in Tab.~\ref{Tab:Pi+}. The complete list of measurements extracted from the spectra is available as supplementary material to this paper online.

For the \pmin\ components comparison is made with Roudjane \etal ~\cite{Roudjane2006} up to $v=18$ and differences are all below 0.5 \wn\ except for the Q(2) transition in the (15,0) band and the Q(2) transition in the (18,0) band. For levels $v \geq 19$ until $v=22$ no previous data exists. Transitions to the \pmin\ levels were also compared with an MQDT calculation~\cite{Glass-Maujean2011} from which we conclude that these calculations are accurate to within 0.5 \wn .  

As discussed in Ref.~\cite{Glass-Maujean2011}, the MQDT calculations disregard the interactions between singly and doubly excited states. These interactions become noticeable at large internuclear distances. These levels are located at high energies and their effect is to push the $D$ state levels downwards. Indeed, the calculated energies are found to be too high by a fraction of a wavenumber. It was shown that the MQDT residuals should asymptotically tend towards the difference
\begin{eqnarray}
\Delta E_{inf} = \dfrac{(R_{H_{2}}-R_{H})}{2n_{d}}
\label{eqn:MQDTResiduals}
\end{eqnarray}
where $R$ is the mass-corrected Rydberg constants and $n_{d}$ is the principal quantum number for the separated atoms; the asymptotic value amounts to 1.6 \wn\ for $n_{d}=3$ in D$_{2}$, which is as much as we find here for the highest $v$ values.

Assignments for the \pplus\ levels were made by comparison with the data of Monfils~\cite{Monfils1965} up to $v=15$. Beyond $v=15$ no measurements exist and so assignments of these transitions were made starting with the emission data of the \pmin\ levels~\cite{Roudjane2006}. Assuming a small $\Lambda$-doublet splitting and correcting for ground state combination differences~\cite{Wolniewicz1995} the R(0) and R(1) lines could be identified using the known Q-lines from the emission study~\cite{Roudjane2006}. For the Q branch transitions beyond $v=19$ again no experimental data exists; we therefore rely on the MQDT calculations to identify these transitions.

\begin{figure}
\centering
\includegraphics[width=1\linewidth]{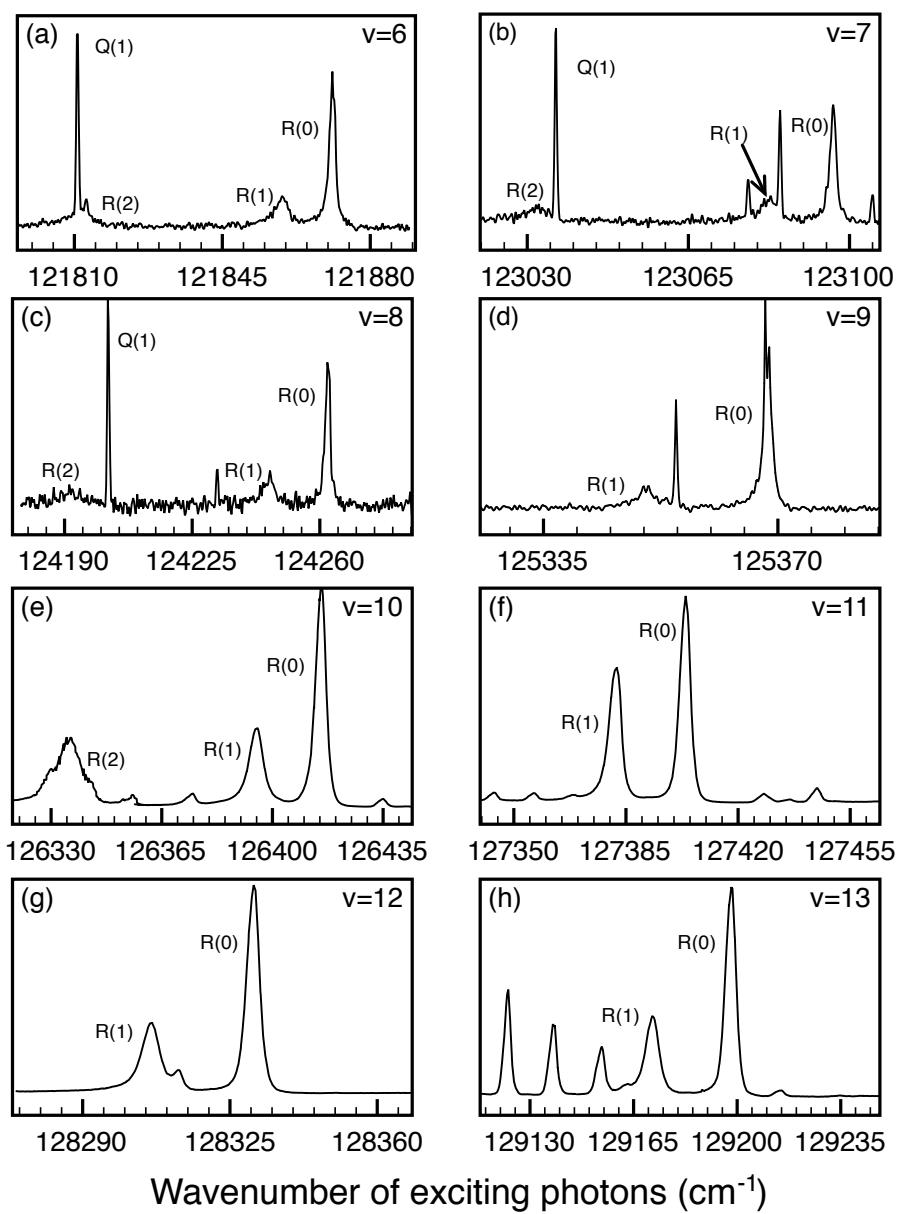}
\caption{Some typical $D-X$ Beutler-Fano resonances. Panels (a)-(d) show the Beutler-Fano resonances in absorption measured with the FT spectrometer at synchrotron SOLEIL. Panels (e)-(h) are recordings of the Ly-$\alpha$ radiation from the dissociated fragments measured at synchrotron BESSY II.}
\label{fig:BeutlerFanoResonances}
\end{figure}

In some cases blended lines were unravelled by deconvolution procedures, as in the example shown in Fig.~\ref{fig:SoleilVsBessy} (b), where the R(0) transition in the (10,0) band is blended with the Q(1) and R(2) transitions from the (4,0) band of the $D''-X$ system~\cite{Takezawa1975}. A similar procedure was followed in the analysis of the R(1) lines in the (11,0),(14,0) and (18,0) bands. Further details of the spectral recordings are illustrated in Fig.~\ref{fig:BeutlerFanoResonances}. Panels (a)-(d) depict absorption measurements of $D-X$ bands (6,0)-(9,0) displaying typical Beutler-Fano resonances.

In the comparison with the previous measurements of Monfils~\cite{Monfils1965}, indicated in Tab.~\ref{Tab:Pi+} for the \pplus components up to $v=15$, differences amount to between 1 and 2 \wn , which is similar to the differences found for the \pmin\ levels when comparing Monfils to Roudjane \etal\ ~\cite{Roudjane2006} and is hence ascribed to the uncertainties in the classical study of Monfils. For the levels $v=16-18$ no previous data exists.

\subsection{Spectroscopic Analysis: BESSY Spectra}
\label{sec:SpectroscopicAnalysis:Bessy}

The line positions of Q branch transitions were obtained from the fluorescence spectrum by means of a Gaussian fit for rotational levels $J=1$ to 4 for vibrations $v=8$ to 23 (at least for $J=1$ and 2).  The uncertainty of 1 \wn\ of these measurements is confirmed by a comparison with the FT measured values with differences less than 0.8 \wn . Further, the agreement with the MQDT calculated values is good with differences amounting to $<1$ \wn\ except for the highest vibration $v=23$.

The R($J$) rotational bands are easily recognized in the Ly-$\alpha$ spectrum; they dominate the spectrum with their intensities and widths (especially for $J>2$) and are not present in the other channels. The vibrational progression can be followed up to the D(1s)+D($n=3$) dissociation limit. Transitions with $v>7$ were analysed by means of fitting to a Fano function convolved with a Gaussian with width ranging from 2-2.5 \wn\ corresponding to the instrument function.  The positions agree to within less than 1 \wn\ when compared to the FT measurements.
Figure~\ref{fig:BeutlerFanoResonances} shows four intervals of the Ly-$\alpha$ spectrum each with $D-X$ resonances for bands (10,0) - (13,0) in panels (e)-(f).

\subsection{$\Lambda$-Doubling}
\label{sec:LambdaDoubling}

\begin{figure}
\begin{center}
\includegraphics[width=1\linewidth]{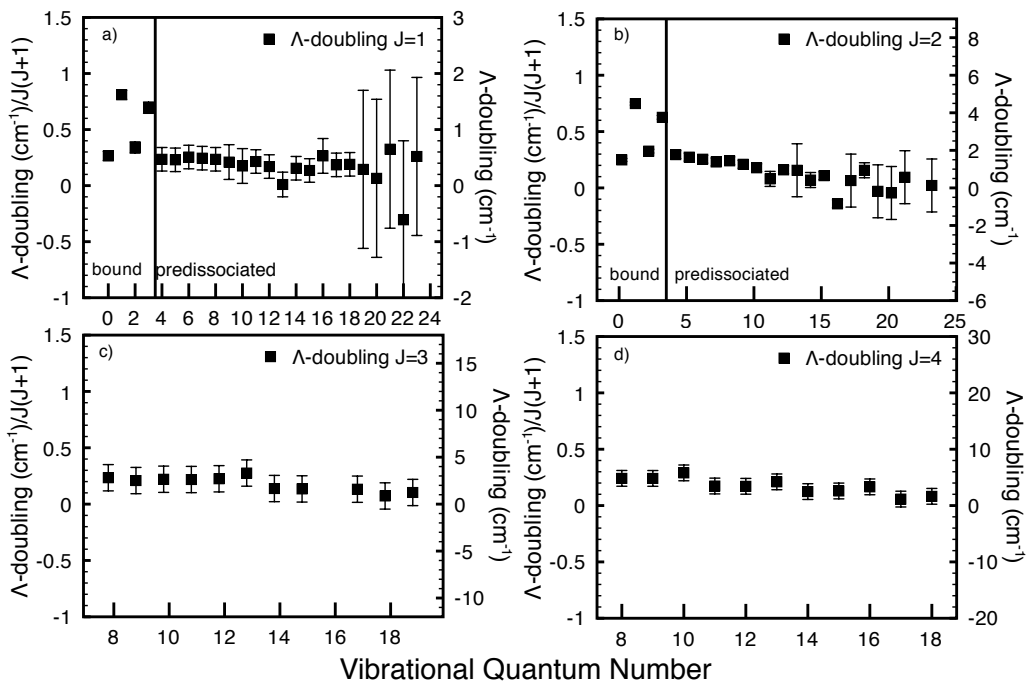}
\caption[]{The extracted $\Lambda$-doublet values (divided through by the factor $J(J+1)$) for the $J=1-4$ levels in the \dstate\ state plotted against vibration. A positive sign indicates that \pplus\ or $\Pi_{e}$ levels are higher than \pmin\ or $\Pi_{f}$ levels. See Tab.~\ref{Tab:Pi-} and~\ref{Tab:Pi+} for details.}
\label{fig:LambdaDoubling}
\end{center}
\end{figure}

The $\Lambda$-doubling is the energy difference between the levels with the same values of $J$ and $v$ but belonging to the two \pplus\ and \pmin\ components. From the listed transition frequencies in the P and R branches, probing \pplus\ levels, and the Q-branch, probing \pmin\ levels the $\Lambda$-doublet splittings in the \dstate\ state can be deduced from the absorption measurements. For this, ground state level intervals of 59.781 \wn\ (between $J=1$ and $J=0$) and 119.285 \wn\ (between $J=2$ and $J=1$) 178.250 \wn\ (between $J=3$ and $J=2$) and 236.398 \wn\ (between $J=4$ and $J=3)$ are used~\cite{Jennings1987,Roudjane2008}. The origin of the $\Lambda$-doubling is found in the couplings which affects the \pplus\ components and not the \pmin\ components: the non-adiabatic couplings with the $\Sigma^{+}$ levels. Such couplings are rotational and their effects are proportional to $J(J+1)$. Resulting values for the $\Lambda$-doublet splittings divided by the factor $J(J+1)$ are displayed in Fig.~\ref{fig:LambdaDoubling} panels (a)-(d); indeed these graphs illustrate the $J(J+1)$ proportionality relation for the $\Lambda$-doubling for all $J$ levels investigated.

As in H$_{2}$ \cite{Dickenson2010a} the bound levels below the second dissociation limit in the \dstate\ system of D$_{2}$ are subject to local perturbations from the last few bound levels of the $B'^{1}\Sigma_{u}^{+}$ state. The interacting $B'$ levels may be located below or above the $D$ state and their energy differences vary from one $D$ level to the other, leading to the erratic behaviour as depicted in Fig.~\ref{fig:LambdaDoubling} (a) and (b) for $v \leq 3$.

Beyond $v=3$ in the region above the $n=2$ dissociation limit, the $\Lambda$-doubling exhibits a smoothly decreasing function with increasing vibration. The values remain positive indicating that \pplus\ or $\Pi_{e}$ levels are higher than \pmin\ or $\Pi_{f}$ levels since they are pushed upward by the $B'$ levels which are at lower energies. 
In fact there are two effects causing the $\Lambda$-doubling to become smaller at higher energies: the wave function overlap between $D$ and $B'$ states is largest at lower energies (and $v$ values), and for high-$v$ levels the energy separation with the states causing the $\Lambda$-doubling gradually increases.


\subsection{ Fano $q$-parameters}
\label{sec:q-parameters}

\begin{figure}
\begin{center}
\includegraphics[width=1\linewidth]{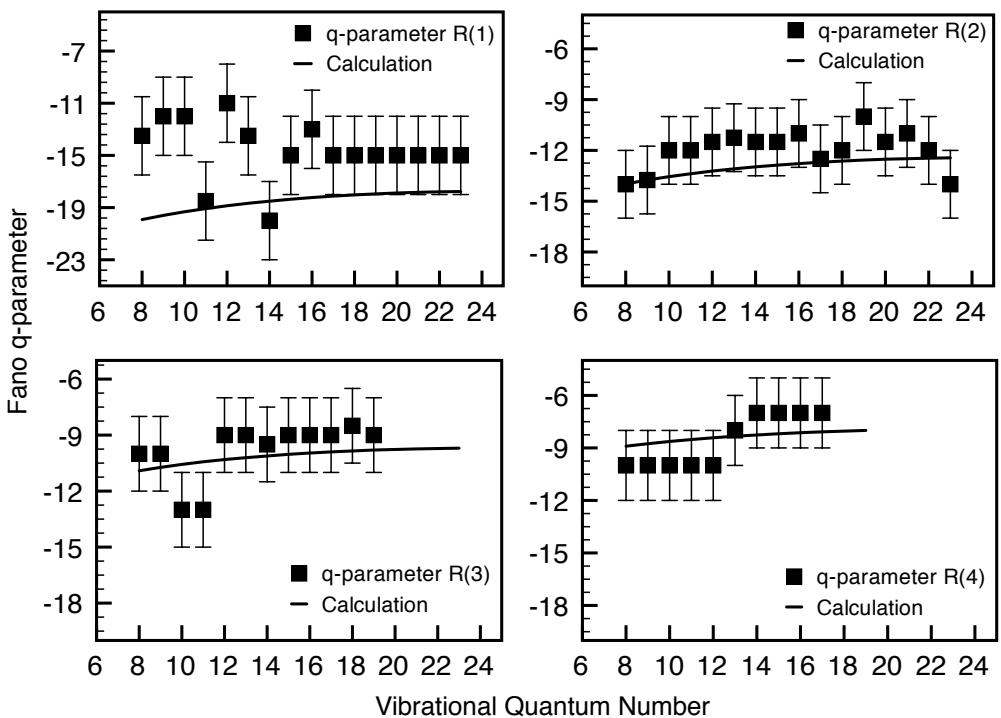}
\caption[]{The extracted Fano $q$-parameters of the $D-X(v',0)$ R(1)-R(4) transitions as a function of $v$ compared with the present calculations. The $q$-parameters were extracted from the Ly-$\alpha$ spectra recorded at BESSY II.}
\label{fig:qParameters}
\end{center}
\end{figure}

The transitions to predissociated \pplus\ levels were fitted with a Fano function. In the R(0) and P(2) lines little asymmetry was detected and the $q$-parameter tended to negative infinity, hence a Lorentzian lineshape.  For the R(1) and R(2) transitions signal to noise issues hampered obtaining a tight constraint on the Fano $q$-parameter from the FT absorption spectrum, notwithstanding the superior resolution. Therefore $q$-parameters are derived from the Ly$-\alpha$ spectrum since signals are recorded on a zero background thereby improving signal to noise ratio and allowing the detection of higher $J$ states. Information on the $q$-parameter for the R(1) - R(4) transitions was obtained and the results are shown in Fig.~\ref{fig:qParameters}. 

Following Eq. (\ref{eqn:qfactor}) and (\ref{eqn:Proportionality}) the $q$-parameters scale like: 
\begin{eqnarray}
q \propto \dfrac{\mu}{\sqrt{J(J+1)}}
\label{eqn:qpropoto}
\end{eqnarray}
The inverse proportionality with the rotational quantum number $J$ follows from the graphs in Fig.~\ref{fig:qParameters}, as is indicated by the comparison with theory yielding good agreement. The mass scaling in Eq.~\ref{eqn:qpropoto} dictates that the $q$-parameters for D$_{2}$ should be twice those of H$_{2}$. The experiments confirm this within the measured uncertainties: the $q$-parameters for the R(1) transitions of H$_{2}$ were found to equal -9$\pm$1~\cite{Glass-Maujean1979a} while for D$_{2}$ a value close to -15$\pm$3 is found. 

\subsection{Predissociated Widths}
\label{sec:Widths}

\begin{landscape}
\begin{figure}
\centering
{\includegraphics[scale=0.6]{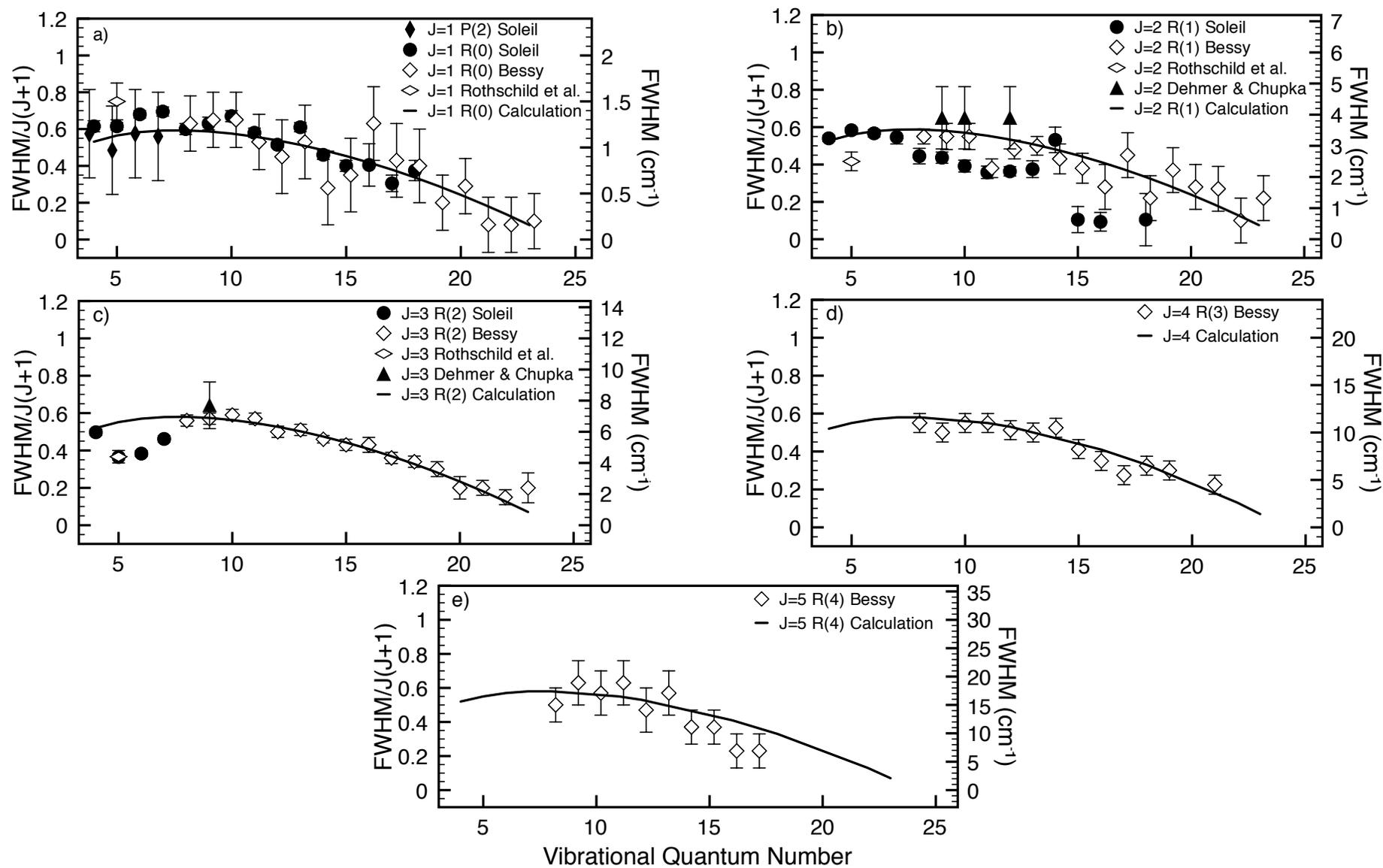}}
\caption[]{The predissociation widths $\Gamma$ for the $J$ = 1-5 levels of the $\mathbf{D^{1}\Pi^{+}_{u}}$ state of D$_{2}$ as extracted from both SOLEIL and BESSY experiments, including comparisons with results from previous experiments~\cite{Rothschild1981,Dehmer1980} and a comparison with the two-state perturbative calculation.}
\label{fig:WidthsAll}
\end{figure}
\end{landscape}

The values for $\Gamma_{v}$ as listed in Tab.~\ref{Tab:Pi+} represent the natural broadening parameters as resulting from a deconvolution procedure, which was explained above. In Fig.~\ref{fig:WidthsAll} panels a) - e) depict the extracted widths from both the FT absorption spectra and the Ly-$\alpha$ spectra for $J=1-5$ along with the corresponding calculated values discussed in section~\ref{sec:Calculation}.  For all $J$ levels $\Gamma$ is a smooth decreasing function of $v$, a consequence of the decreasing vibrational overlap between the $D$ state and the $B'$ continuum.

There is good agreement with previously measured values of Rothschild \etal\ \cite{Rothschild1981} and Dehmer and Chupka~\cite{Dehmer1980}. The values extracted from the FT absorption spectrum and those from the Ly-$\alpha$ spectrum agree well, thereby confirming experimental consistency. Further the widths for rotational levels $J=4$ and 5 extracted from the Ly-$\alpha$ spectrum are shown in~\ref{fig:WidthsAll} d) and e) respectively. There are no previous measurements for these transitions.

The agreement between the measured values and the perturbation calculation is very good for all $J$ levels, thereby confirming the scaling in Eq.(\ref{eqn:Width}) and showing that the predissociation is dominated by the $B'-D$ coupling.

\subsection{Line Intensities}
\label{sec:Intensities}
We have extracted Einstein coefficients corresponding to the observed $D-X$ transitions using the relation
\begin{eqnarray}
\sigma = \int (\sigma(\lambda)-\sigma_{cont})d\lambda = \dfrac{\lambda^{4}}{8\pi c}A_{v',v''}N_{J''}h(J',J'')
\label{eqn:IntegratedCrossSection}
\end{eqnarray}
Here the integrated absorption cross section of the discrete state (line intensity), $\sigma$, is obtained by integrating the measured absorption cross section, $\sigma(\lambda)$, over the profile of a given line, taking the continuum baseline as origin ($\sigma_{cont}$). $\lambda$ is the wavelength, $N_{J''}$ is the fraction of molecules in the rotational state $J''$ and $h$($J'$,$J''$) is the rotational H\"{o}nl-London factor: equal to 1 for a Q($J''$) transition and $(J''+2)/(2J''+1)$ for a R($J''$) transition. The measured $\sigma$ values are available as supplementary material to this paper online.

\subsubsection{States of \pmin\ Symmetry}

\begin{figure}
\centering
\includegraphics[width=1\linewidth]{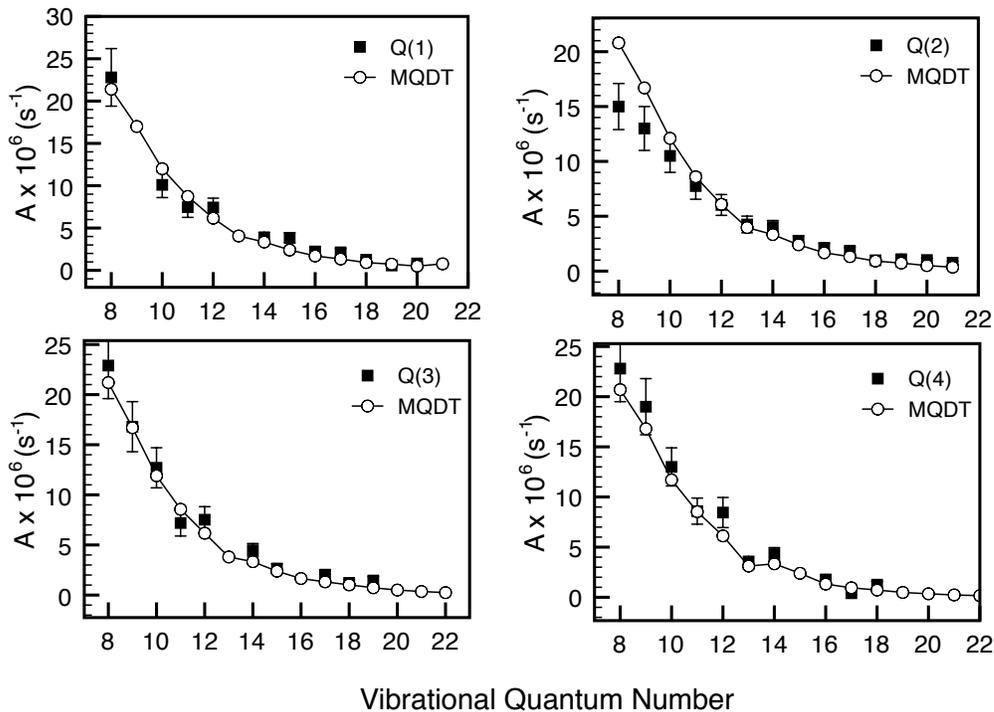}
\caption{The intensities of the Q-branch transitions extracted from the fluorescence excitation spectrum (filled squares). The intensities are compared to an MQDT calculation~\cite{Glass-Maujean2011} (open circles).}
\label{fig:IntensitiesQ}
\end{figure}

The area of each Q-branch transition, as observed in the absorption or difference spectrum measured at BESSY II, was evaluated by means of a Gaussian fit.  Figure~\ref{fig:IntensitiesQ} shows measured A$_{v',v''=0}$ values for the various \pmin\ $J$ (excited state $J$) levels of the $D$ state. The observed values agree very well with the values calculated in the MQDT framework, which are also in agreement with results from coupled differential equations~\cite{Roudjane2006}, except for Q(2) $v=8$ and 9 lines. These lines are very intense and are partly saturated.

\subsubsection{States of \pplus\ Symmetry}

\begin{figure}
\centering
\includegraphics[width=1\linewidth]{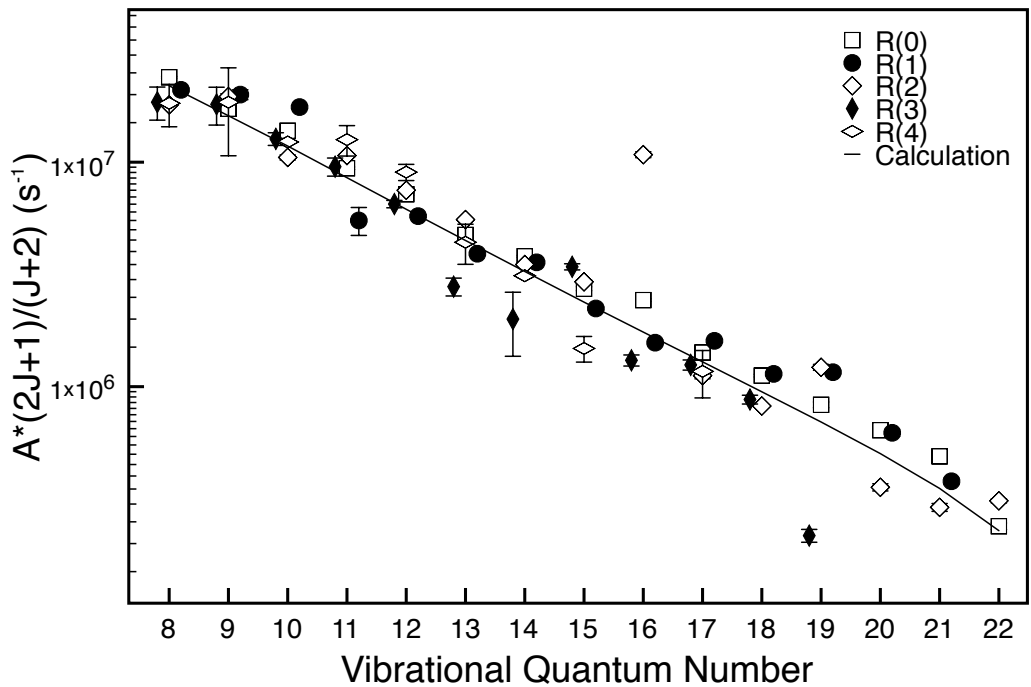}
\caption{The intensities for the R-branch transitions extracted from the Ly-$\alpha$ spectrum. The calculation was made using the perturbative model based on a single bound state interacting with a continuum.}
\label{fig:IntensitiesR}
\end{figure}

As the dissociation of the $D^{1}\Pi_{u}^{+}$ levels ($v>$3) is fast compared to the other decay channels (ionization and 
fluorescence), their dissociation yield is unity, the integral in Eq.~\ref{eqn:IntegratedCrossSection} can be evaluated from the Ly-$\alpha$ excitation spectrum. 
Figure~\ref{fig:IntensitiesR} shows the measured $A_{v',v''=0}$ values for the various $J$ levels of the $D$ state, compared to the 
perturbative calculations. The observed values for the $D^{1}\Pi^{+}_{u}$ levels agree to within 30$\%$ (note the logarithmic scale) with the calculated values except for: the R(3) $v=13, 14$ and 20 and the R(4) $v=15$ lines for which deviations by up to a factor of $2.5$ are observed, indicative of local perturbations associated with channels not accounted for. The present perturbative model calculation does not include non-adiabatic couplings and the ionization channel is also neglected. 

\section{Conclusion}

The $D^{1}\Pi_{u}$ state in molecular hydrogen is a benchmark system for the study of the predissociation phenomenon. The present analysis extends the study to D$_{2}$ providing information on mass-dependent effects. Two independent and complementary experimental setups at advanced third generation synchrotron facilities have been employed for a full characterization of the \dstate\ state in D$_{2}$, in terms of spectroscopic determination of level energies, determination of absolute cross sections, as well as the dynamical decay rates of the excited states. The high-resolution VUV Fourier-Transform-spectrometer instrument at SOLEIL provides line positions for transitions in the $D-X$ system at high accuracy, in particular for the narrow Q-branch lines. An exerimental setup at the 10 m normal-incidence beamline at BESSY II equipped
with four different detection instruments (for absorption, molecular fluorescence, dissociation via Ly-$\alpha$ fluorescence, and ionization) operating simultaneously, yield accurate information on absolute cross sections, for absorption and for the rates of decay between competing channels.

In fact two different predissociation mechanisms play a role in the decay of the \dstate\ state. The \pplus\ components of the \dstate\ state strongly couple via a heterogeneous (rotational) interaction to the bound levels and the continuum of the $B'^{1}\Sigma_{u}^{+}$ state. The observational data on \pplus\ levels are modelled based on a two-state perturbation calculation describing this interaction. This simplified model is able to explain the extensive new data set on absorption cross sections in the $D-X$ system, as well as line broadening parameters and Fano $q$-parameters even in the energy range above the ionization potential.

For the \pmin\ components of the \dstate\ state interaction with the $B'^{1}\Sigma_{u}^{+}$ state is symmetry forbidden. Weak predissociation, mainly due to coupling with the lower lying $C^{1}\Pi_{u}$ state, is the driving mechanism for predissociation. In view of the weakness of this coupling, Q-branch lines are narrow, and the subtle effects of coupling to manifolds of (-) symmetry levels must be accounted for. This is accomplished in non-adiabatic MQDT framework. The extended observational data on $D^{1}\Pi_{u}^{-}$ states (both line positions and absorption cross sections) are found to be in excellent agreement with these fully \emph{ab initio} calculations.

\section{Supplementary Material}
An electronic database containing all measured transition frequencies, level energies, intensities, $\Gamma$ values and $q$-parameters for the \pplus\ parity components and where applicable for \pmin\ parity components is available as supplementary material to this paper online.

\section{Acknowledgements}

GDD is grateful to the general SOLEIL staff for the hospitality and for running the facility. This work was supported by the Netherlands Foundation for Fundamental Research of Matter (FOM). The EU provided financial support through the transnational funding scheme. MGM acknowledges the Helmholtz-Zentrum Berlin electron storage ring BESSY II for provision of synchrotron radiation at beam line U125/2-10m-NIM and would like to thank G Reichardt and P. Baumg\"{a}rtel for assistance. The research leading to these results have received funding from the European Community's Seventh Framework Programme (FP7/2007-2013) under Grant no 226716. W.-\"{U} L. T-B acknowledges supports from the ANR (France) under contract N$^{\circ}$ 09-BLAN-020901 and from the CNRS/INSU (France) PNPS programme. AK acknowledges support from the Otto Braun Fond.

\clearpage

\clearpage

\begin{longtable}{l r r c l r r c }
\caption[]{Transition frequencies of Q branch lines probing levels of $\Pi^{-}$ symmetry for the $D^{1}\Pi_{u}$ - \xstate ($v'$,0) system of D$_{2}$. The full list of measurements is available as supplementary material to this paper online. $\Delta$ represent deviations in \wn\ calculated from level energies from Roudjane \etal~\cite{Roudjane2006}. $\Delta^{MQDT}$ is a comparison between present values and an MQDT calculation. Lines marked with a superscript $b$ are blended, those marked with a superscript $s$ were slightly saturated and those marked with an $f$ are extracted from the fluorescence spectrum recorded at BESSY II. All values in \wn .}
\label{Tab:Pi-}\\
\colrule
\multicolumn{2}{c}{Transition Frequency} & \multicolumn{1}{c}{$\Delta$}  & \multicolumn{1}{c}{$\Delta^{MQDT}$}&\multicolumn{2}{c}{Transition Frequency} & \multicolumn{1}{c}{$\Delta$}  & \multicolumn{1}{c}{$\Delta^{MQDT}$}    \\ 
\colrule
\endfirsthead

\colrule
\multicolumn{2}{c}{Transition Frequency} & \multicolumn{1}{c}{$\Delta$}  & \multicolumn{1}{c}{$\Delta^{MQDT}$}&\multicolumn{2}{c}{Transition Frequency} & \multicolumn{1}{c}{$\Delta$}  & \multicolumn{1}{c}{$\Delta^{MQDT}$}   \\ 
\colrule	
\endhead

\colrule
\multicolumn{8}{c}{{continued on next page}} \\ 
\endfoot

\colrule
\endlastfoot

$D-X(0,0)$& & & &$D-X(1,0)$ & & & \\
Q(1)&113162.66(6) &-0.21 &-0.31 & Q(1)&114763.71(6) &0.07 &-0.26 \\
Q(2)&113102.89(6) & -0.14&-0.24 & Q(2)&114701.71(6) &0.01 &-0.22\\
& & & &Q(3)&114609.09(6) &-0.01 &-0.23  \\
\\
$D-X(2,0)$& & & &$D-X(3,0)$ & & & \\
Q(1)&116299.09(6) &-0.13 &-0.22 & Q(1)&117770.29(6) &-0.14 &-0.19  \\
Q(2)&116234.93(6) & -0.12&-0.20 & Q(2)&117704.02(6) &-0.12&-0.19\\
Q(3)&116139.09(6) &-0.23 &-0.21 & Q(3)&117605.01(6) &-0.13&-0.20 \\
 & & & & Q(4) &117473.39(6)&-0.44 &-0.59 \\
\\
$D-X(4,0)$& & & &$D-X(5,0)$ & & & \\
Q(1)&119178.60(6)&-0.20&-0.15 & Q(1)&120525.13(6) &-0.14&-0.09 \\
Q(2)&119110.26(6) & -0.16&-0.16 & Q(2)&120454.74(6) &-0.21&-0.13\\
Q(3)&119008.13(6) &0.02 &-0.19 & Q(3)&120349.56(6) &-0.34 &-0.15\\
Q(4)&118872.79(6)&-0.01&-0.17& Q(4)&120210.15(6)&0.00&-0.14 \\
\\
$D-X(6,0)$& & & &$D-X(7,0)$ & & & \\
Q(1)&121810.78(6)&0.03 &-0.04 & Q(1)&123036.22(6) &0.02 &-0.01\\
Q(2)&121738.35(6)& -0.04&-0.12& Q(2)&122961.79(6)&-0.01 &-0.10\\
Q(3)&121630.17(6) &0.10 &-0.12& Q(3)&122850.63(6)&0.06 &-0.10\\
Q(4)&121486.72(6)&0.10&-0.14& & & & \\ 
\\
$D-X(8,0)$& & & &$D-X(9,0)$ & & & \\
Q(1)&124201.97(6)&0.08&-0.03& Q(1)$ ^{s} $&125308.28(8)&-0.09&0.02  \\
Q(2)&124125.59(6)&0.16&-0.10&Q(2) &125229.95(6)&0.07&0.00\\
Q(3)&124011.52(6)&0.17&0.05&Q(3)$^{f}$ & 125113.3(10)&0.54&0.36\\
Q(4)$^{f}$&123860.3(10)&-0.15&0.11&Q(4)$^{f}$ & 124957.8(10)&0.37 &0.27 \\
\\
$D-X(10,0)$& & & &$D-X(11,0)$ & & & \\
Q(1)$ ^{s} $&126355.38(8)&-0.07&0.22& Q(1)&127342.94(6)&0.05 &0.15 \\
Q(2)&126275.03(6)& -0.22&0.13& Q(2)&127260.60(6)&-0.22&0.14\\
Q(3)$^{f}$&126155.2(10)&0.11&0.22&Q(3)$^{f}$ & 127137.6(10)&0.23&0.08 \\
Q(4)$^{f}$&125995.7(10)&-0.17&0.05&Q(4)$^{f}$ & 126974.5(10)&0.40&0.26 \\
\\
$D-X(12,0)$& & & &$D-X(13,0)$ & & & \\
Q(1)&128270.57(6)&-0.05&0.32&Q(1)$^{s} $&129137.85(8)&0.23&0.62\\
Q(2)&128186.22(6)& -0.03&0.16&Q(2)$^{f}$&129051.5(10)&0.30&0.30\\
Q(3)&128060.22(6)&0.13&0.23&Q(3)$^{f}$& 128921.1(10)&-1.19&-0.70\\
Q(4)$^{f}$&127892.8(10)&-0.02&0.02&Q(4)$^{f}$ & 128751.5(10)&0.87&0.88\\
\\
$D-X(14,0)$& & & &$D-X(15,0)$ & & & \\
Q(1)&129942.79(6)&-0.24&0.43&Q(1)&130685.13(6)&-0.11&0.51\\
Q(2)&129854.29(6)&0.06&0.35&Q(2)&130594.46(6)&-0.74&0.47\\
Q(3)$^{f}$&129722.0(10)&0.08&0.25&Q(3)$^{f}$ & 130459.0(10)&0.16&0.45\\
Q(4)$^{f}$&129546.6(10)&-0.12&0.30&Q(4)$^{f}$ & 130279.5(10)&0.38&0.77\\
\\
$D-X(16,0)$& & & &$D-X(17,0)$ & & & \\
Q(1)&131362.81(6)&-0.18&0.61&Q(1)&131973.99(6)&0.15&0.86  \\
Q(2)&131269.92(6)&-0.29&0.49&Q(2)$^{f}$&131878.7(10)&-0.11&0.61\\
 & & & &Q(3)$^{f}$& 131736.0(10)&-0.05 &0.21\\
Q(4)$^{f}$&130946.0(10)&-2.89&-0.41&Q(4)$^{f}$& 131547.8(10)&0.13&0.83 \\
\\
$D-X(18,0)$& & & &$D-X(19,0)$ & & & \\
Q(1)&132516.13(6)&-0.20&0.86&Q(1)&132986.74(6)&---&0.94 \\
Q(2)&132418.53(6)& 0.79&0.82&Q(2)&132886.53(6)&0.54&0.75\\
Q(3)$^{f}$&132272.5(10)&-0.23&0.73&Q(3)$^{f}$& 132736.8(10)&-0.17&0.93\\
Q(4)$^{f}$&132078.8(10)&0.21&0.79& & & & \\
\\
$D-X(20,0)$& & & &$D-X(21,0)$ & & & \\
Q(1)&133383.30(6)&---&1.06& Q(1)&133702.83(6)&---&1.02  \\
Q(2)&133280.35(6)&---&0.86&Q(2)$^{f}$ &133597.0(10)&---&0.91\\
\\
$D-X(22,0)$& & & &$D-X(23,0)$ & & & \\
Q(1)&133942.65(6)&---&1.22&Q(1)$^{f}$&134098.7(10) &---&2.30  \\
Q(2)$^{f}$&133832.9(10)&---&0.27&Q(2)$^{f}$&133987.3(10)&---&3.30\\

\end{longtable}

\begin{longtable}{l r @{.} l r r l l @{.} l r r}
\caption[]{Transition frequencies of R and P branch lines probing levels of $\Pi^{+}$ symmetry for the $D^{1}\Pi_{u}$ - \xstate ($v'$,0) system of D$_{2}$. A full list of the measurements is available as supplementary material to this paper online. $\Delta$ represents deviations in \wn\ calculated from level energies reported by Monfils~\cite{Monfils1965} for $v=4-15$ and those reported by Roudjane \etal\ ~\cite{Roudjane2006} for $v=0-3$. $\Gamma$ represents the deconvolved values for the predissociated widths. Lines marked with a $b$ were blended, those marked with an $s$ were slightly saturated and those marked with an $l$ are extracted from the Ly-$\alpha$ spectrum. All values in \wn\ and uncertainties in the line positions are given in brackets.}
\label{Tab:Pi+}\\
\colrule
\multicolumn{3}{c}{Transition Frequency} & \multicolumn{1}{c}{$\Delta$}  & \multicolumn{1}{c}{$\Gamma$}&\multicolumn{3}{c}{Transition Frequency} & \multicolumn{1}{c}{$\Delta$}  & \multicolumn{1}{c}{$\Gamma$}    \\ 
\colrule
\endfirsthead

\colrule
\multicolumn{3}{c}{Transition Frequency} & \multicolumn{1}{c}{$\Delta$}  & \multicolumn{1}{c}{$\Gamma$}&\multicolumn{3}{c}{Transition Frequency} & \multicolumn{1}{c}{$\Delta$}  & \multicolumn{1}{c}{$\Gamma$}   \\ 
\colrule	
\endhead

\colrule
\multicolumn{8}{c}{{continued on next page}} \\ 
\endfoot

\colrule
\endlastfoot
\multicolumn{5}{l}{$D-X (0,0)$}&\multicolumn{3}{l}{$D-X (1,0)$}\\
R(0)&113222&97(6)&-0.09&---&R(0)&114825&11(6)&-0.24&---\\
R(1)&113223&66(8)&-0.07&---&R(1)&114825&48(8)&0.01&---\\
R(2)&113194&64(6)&-0.03&---&R(2) &114795&55(6)&0.04 &---\\
P(2)&113043&95(6)&-0.10&---&P(2) &114646&05(8)&-0.23&---\\
\\
\multicolumn{5}{l}{$D-X (2,0)$}&\multicolumn{3}{l}{$D-X (3,0)$} \\
R(0)&116359&56(6)&-0.33&---&R(0)&117831&47(6)&0.11&---\\
R(1)&116356&18(6)&-0.09&---&R(1)&117827&06(6)&-0.10 &---\\
R(2)&116321&03(6)&0.17&---&R(2) &117789&87(6)&0.16 &---\\
P(2)&116180&48(6)&-0.33&---&P(2) &117652&39(6)&0.10 &---\\
\\
\multicolumn{5}{l}{$D-X (4,0)$}&\multicolumn{3}{l}{$D-X (5,0)$}  \\
R(0)&119238&85(20)&1.86&1.2& R(0)&120585&38(20)&1.58&1.2\\
R(1)&119231&32(20)&1.40&3.2&R(1) &120575&66(20)&1.17 &3.5\\
R(2)&119189&79(40)&-0.03&6.0&R(2) &120531&51(40)&0.61&4.4\\
P(2)&119059&91(20)&1.98&1.2&P(2) &120406&60(20)&-1.87&1.0\\
\\
\multicolumn{5}{l}{$D-X (6,0)$}&\multicolumn{3}{l}{$D-X (7,0)$}\\
R(0)&121871&07(20)&-0.97&1.4&R(0)&123096&49(20)&1.14&1.4\\
R(1)&121859&16(20)&0.74&3.4&R(1) &123082&46(20)&0.95&3.3\\
R(2)&121812&08(40)&2.03&4.3&R(2) &123032&23(40)&0.49&5.5\\
P(2)&121692&18(20)&-0.8&1.2&P(2) &122917&46(20)&1.18&1.1\\
\\
\multicolumn{5}{l}{$D-X (8,0)$}&\multicolumn{3}{l}{$D-X (9,0)$} \\
R(0)&124262&22(20)&3.32&0.8&R(0)$ ^{s} $&125368&48(40)&1.54&1.3\\
R(1)&124246&32(20)&0.72&2.7&R(1) &125350&50(20)&-0.07&2.6\\
R(2)$^{l}$&124192&5(10)&1.18&6.7&R(2)$^{l}$&125294&0(10)&0.6&6.8\\
R(3)$^{l}$&124101&5(10)&-0.11&11&R(3)$^{l}$&125199&0(10)&2.20&10\\
R(4)$^{l}$&123972&6(10)&1.01&15&R(4)$^{l}$&125063&3(10)&---&19\\
\\
\multicolumn{5}{l}{$D-X (10,0)$}&\multicolumn{3}{l}{$D-X (11,0)$} \\
R(0)$^{b}$&126415&51(40)&0.6&1.3&R(0)&127403&15(20)& 1.07&1.2\\
R(1)&126395&39(20)&0.00&2.4&R(1)$^{b}$ &127380&37(40)&1.27&2.2\\
R(2)$^{l}$&126336&0(10)&-0.82&7.1&R(2)$^{l}$ &127318&4(10)&1.59&6.8\\
R(3)$^{l}$&126237&9(10)&0.53&11&R(3)$^{l}$&127214&4(10)&0.15&11\\
R(4)$^{l}$&126098&6(10)&-0.37&17&R(4)$^{l}$ &127074&7(10)&--- &19\\
\\
\multicolumn{5}{l}{$D-X (12,0)$}&\multicolumn{3}{l}{$D-X (13,0)$}\\
R(0)&128330&70(20)&0.19&1.0&R(0)&129197&66(20)& 1.38&1.2\\
R(1)&128306&46(20)&-0.14&2.2&R(1) &129171&26(20)&-0.20&2.3\\
R(2)$^{l}$&128240&7(10)&-0.54&6.1&R(2)$^{l}$&129102&6(10)&1.29&5.9\\
R(3)$^{l}$&128132&6(10)&-2.72&10&R(3)$^{l}$&128992&1(10)&1.60&10\\
R(4)$^{l}$&127982&8(10)&-5.43&14&R(4)$^{l}$ &128835&3(10)&-3.20&17\\
\\
\multicolumn{5}{l}{$D-X (14,0)$}&\multicolumn{3}{l}{$D-X (15,0)$} \\
R(0)&130002&88(20)&-0.30&0.9&R(0)&130745&18(20)&0.94&0.8\\
R(1)$^{b}$&129974&0(1)&0.70&3.2&R(1) &130714&40(20)&2.95&0.6\\
R(2)$^{l}$&129102&6(10)&2.57&5.5&R(2)$^{l}$&130638&8(10)&0.42&5.2\\
R(3)$^{l}$&129785&4(10)&1.97&10&R(3)$^{l}$&130518&5(10)&0.19&8.3\\
R(4)$^{l}$&129625&7(10)&5.75&11&R(4)$^{l}$&130352&6(10)&9.27&11\\
\\
\multicolumn{5}{l}{$D-X (16,0)$}&\multicolumn{3}{l}{$D-X (17,0)$} \\
R(0)$^{b}$&131423&12(40)&---&0.8&R(0)&132034&13(20)&---&0.6\\
R(1)&131388&35(20)&---&0.6&R(1)$^{l}$&131998&4(10)&---&2.7 \\
R(2)$^{l}$&131311&1(10)&---&5.1&R(2)$^{l}$ &131915&8(10)&---&4.4\\
R(3)$^{l}$&131185&7(10)&---&7.0&R(3)$^{l}$ &131785&3(10)&--- &5.5\\
\multicolumn{5}{c}{}&R(4)$^{l}$ &131608&4(10)&--- &7.0\\
\\
\multicolumn{5}{l}{$D-X (18,0)$}&\multicolumn{3}{l}{$D-X (19,0)$} \\
R(0)&132576&30(20)&---&0.8&R(0)$^{l}$&133046&7(10)&--- & 0.4\\
R(1)$^{b}$&132538&76(40)&---&0.6&R(1)$^{l}$&133005&8(10)&---&2.2\\
R(2)$^{l}$&132451&6(10)&---&4.1&R(2)$^{l}$ &132916&2(10)&---&3.7\\
R(3)$^{l}$&132316&9(10)&---&6.5&R(3)$^{l}$ &132775&0(10)&--- &6.0\\
R(4)$^{lb}$&132133&6(10)&---&---&\multicolumn{5}{c}{}\\
\\
\multicolumn{5}{l}{$D-X (20,0)$}&\multicolumn{3}{l}{$D-X (21,0)$} \\
R(0)$^{l}$&133442&9(10)&---&0.6&R(0)$^{l}$&133763&0(10)&---&0.2\\
R(1)$^{l}$&133400&0(10)&---&1.7&R(1)$^{l}$&133716&8(10)&---&1.6\\
R(2)$^{lb}$&133305&5(10)&---&---&R(2)$^{l}$&133617&8(10)&---&4.5\\
\multicolumn{5}{c}{}&R(3)$^{lb}$&133478&4(10)&--- &---\\
\\
\multicolumn{5}{l}{$D-X (22,0)$}&\multicolumn{3}{l}{$D-X (23,0)$} \\
R(0)$^{l}$&134002&6(10)&---&0.2&R(0)$^{l}$&134159&0(10)& ---&0.2\\
\multicolumn{5}{c}{}&R(1)$^{l}$&134106&7(10)&---&1.3\\
R(2)$^{l}$&133733&9(10)&---&1.9&\multicolumn{5}{c}{}\\

\end{longtable}

\end{document}